\pdfoutput=1 
\documentclass[useAMS,usenatbib]{mn2e}
\usepackage{pdf14,graphicx,rotate,url,mathptmx,lscape,times}
\def\gtsim{\mathrel{\hbox{\rlap{\hbox{\lower4pt\hbox{$\sim$}}}\hbox{$>$}}}}
\def\lesssim{\mathrel{\hbox{\rlap{\hbox{\lower4pt\hbox{$\sim$}}}\hbox{$<$}}}}

\def\A{{\rm\thinspace \AA}}
\def\cm{{\rm\thinspace cm}}

\def\erg{{\rm\thinspace erg}}

\def\K{{\rm\thinspace K}}

\def\km{{\rm\thinspace km}}

\def\s{{\rm\thinspace s}}
\def\ps{{\rm\thinspace s^{-1}}}

\def\ergpccmps{\hbox{$\erg\cm^{-3}\s^{-1}\,$}}
\def\ergpscmps{\hbox{$\erg\cm^{-2}\s^{-1}\,$}}

\def\ergpscmpspa{\hbox{$\erg\cm^{-2}\s^{-1}\A^{-1}\,$}}

\def\kmps{\hbox{$\km\ps\,$}}

\def\pscm{\hbox{$\cm^{-2}\,$}}

\def\pccm{\hbox{$\cm^{-3}\,$}}
\def\pscm{\hbox{$\cm^{-2}\,$}}

\def\pccmK{\hbox{$\cm^{-3}\K$}}

\DeclareMathAlphabet{\vib}{OML}{cmm}{m}{it}

\title[Two-Photon Emission from a Filament in NGC~1275]{Hydrogen Two-Photon
Continuum Emission from the Horseshoe Filament in NGC~1275}

\author[R.M. Johnstone et al.]
       {\parbox[]{10.0in}
       {R.M. Johnstone$^1$\thanks{E-mail: rmj@ast.cam.ac.uk}, 
        R.E.A. Canning$^{1}$, A.C. Fabian$^{1}$, G.J. Ferland$^2$, M.
Lykins$^2$,\\ R.L. Porter$^3$, P.A.M. van Hoof$^4$ and R.J.R Williams$^5$\\
        \footnotesize
        $^1$Institute of Astronomy, University of Cambridge, Madingley Road,
Cambridge CB3 0HA\\
        $^2$Department of Physics, University of Kentucky, Lexington, KY 40506,
USA\\
        $^3$Department of Physics and Astronomy and Center for Simulational
Physics, The University of Georgia, Athens, GA 30602-2451, USA\\
        $^4$Royal Observatory of Belgium, Ringlaan 3, 1180 Brussels, Belgium\\
        $^5$AWE plc, Aldermaston, Reading RG7 4PR
}}

\date{
      Received }

\pagerange{\pageref{firstpage}--\pageref{lastpage}}
\pubyear{}

\begin{document}

\maketitle

\label{firstpage}

\begin{abstract}
\noindent
Far ultraviolet emission has been detected from a knot of H$\alpha$ emission in
the Horseshoe filament, far out in the NGC~1275 nebula. The flux detected
relative to the brightness of the H$\alpha$ line in the same spatial region  is
very close to that expected from Hydrogen two-photon continuum emission in the
particle heating model of \citet{Ferlandetal09} if reddening internal to the
filaments is taken into account. We find no need to invoke other sources of far
ultraviolet emission such as hot stars or emission lines from CIV in
intermediate temperature gas to explain these data. 

\end{abstract}


\section{Introduction}
\label{intro}
Many central galaxies in cool core clusters exhibit an extensive optical
emission-line nebula. The best observed example of such a system is in NGC~1275
which lies at the centre of the Perseus Cluster. The optical emission-line
spectrum shows a characteristic low-ionization state similar to that seen in Low
Ionization Nuclear Emssion-Line Regions (LINERs); many authors have sought to
understand the excitation and ionization mechanism of this gas. The line
spectrum is rich and there are many potential diagnostics present.

Recently, the field has become confused because, in the nearby systems that we
can observe in detail (eg NGC~4696 in the centaurus Cluster and NGC~1275 in the
Perseus Cluster), it is clear that different processes are at work in different
spatial regions. Depending on the redshift of the source, the instrument and the
exposure times being used different parts of the nebula will be detected. For
example, in early works such as \citet{Johnstoneetal87} which used smaller
telescopes and relatively short exposure times the emission lines were detected
mainly from the very central part of the galaxy. There, it was found that an
excess of blue light over that expected from a normal elliptical galaxy
correlated quantitatively with the luminosity in H$\alpha$, strongly suggesting
the presence of star formation.  Star formation is also clearly at work in some
regions which are much further out (e.g. \citealt{Canningetal10}).

Most central cluster galaxies are host to a radio source, indicating the
presence of an active nucleus. Although in most cases the 
accretion luminosity from the black hole is many orders of magnitude lower than
expected from accretion at the Bondi rate (\citealt{Dimatteoetal01}) there are
some sources which are clearly dominated by lines from the active nucleus in
their central regions (eg broad lines in Abell~1068, {\citealt{Hatchetal07};
strong [OIII]$\lambda5007$  emission in NGC~1275, \citealt{JohnstoneFabian88};
strong [OIII]$\lambda5007$ emission, Abell~3581 \citealt{Farageetal12}).

In a series of papers starting with \citet{Johnstoneetal07} we are focussing on
the emission line spectrum from more typical regions of the nebulae, far away
from the complex regions near the centres of the galaxies. Although the surface
brightness is lower, the expectation is that the physical processes at work are
simpler and therefore easier to understand. The Horseshoe filament in NGC~1275,
or more particularly the bright knot in that region known as Region 11 in the
notation of \citet{Conseliceetal01}, is at a radial distance of 22 kpc (62
arcsec) from the centre of the galaxy and has an emission-line spectrum
characteristic of large areas of the emission line nebulae
(\citealt{Hatchetal06}). There are several key diagnostics which set these
regions apart from most other astrophysical emission line regions: First, the
molecular hydrogen lines are very strong relative to the Balmer or Paschen lines
of hydrogen, and certainly much stronger than can be produced in a
photo-dissociation region such as the 
Orion bar (\citealt{Hatchetal05}; \citealt{Ferlandetal08}). Second, the
forbidden [NI] doublet at $\lambda5199$\AA\ is unusually strong, being typically
$0.25\times$H$\beta$ (\citealt{Ferlandetal09}). Third, the [NeIII] lines, both
in the optical at $\lambda 3869$\AA, and in the mid-infrared at
$\lambda15.55\mu$m are stronger than expected while the
[OIII]$\lambda\lambda4959,5007$ lines are typically very weak
{(\citealt{Ferlandetal09})}.

\section{Particle Heating}
\label{pheat}
\citet{Ferlandetal09} used the code {\sc cloudy} to calculate the \emph
{constant pressure} emission-line spectrum from Region 11 of the NGC~1275 nebula
in which the excitation and ionization of the filaments was driven purely by
collisions with ionizing particles. (The background level of starlight relevant
to the galaxy environment of the nebula was included for completeness but had
minimal effect due to its low intensity and an assumed self-shielding equivalent
Hydrogen column density of $10^{21}\pscm$.) That model was able to reproduce the
strengths of the commonly observed emission lines (including the key diagnostics
mentioned above) within a factor of two, using only one free parameter, a
power-law slope constraining the relative amounts of hot and cold gas.

In the physical model, the target cool gas exists at a range of densities and
temperatures that are in pressure equilibrium with the surrounding X-ray
emitting intracluster medium gas. Particle ionization can occur at a range of
rates. To obtain the predicted spectrum, a grid of models was calculated
covering a range of target gas densities and a range of ionizing particle
fluxes. A weighted sum of these models was taken along a locus of constant
pressure ($nT=10^{6.5}\pccmK$) matching that of the intracluster medium at
Region 11. The weighting factors were constrained so that the cumulative volume
filling factor at each temperature varied as a power-law in density of index
$\alpha$. $\alpha$ was the only free parameter in the model; it was set to
reproduced the observed ratio of H$_2$~12.28$\mu$m / H$\alpha$ = 0.03.

In \citet{Ferlandetal09} this model is referred to in some places as the `cosmic
rays' model to distinguish it from a separate model that was considered in which
only thermal heat was injected in to the gas. In practice, all that is required
in the particle heating model are particles with kinetic energies several times
the ionization energy of Hydrogen; relativistic particles or actual cosmic rays
are not required. Observations of NGC~1275 (\citealt{Fabianetal03}) and NGC~4696
(\citealt{Crawfordetal05}) have shown a detailed spatial correspondence between
the coolest observed X-ray emitting gas and H$\alpha$ emitting filaments leading
\citet{Fabianetal11} to discuss the possibility that the particles involved in
exciting the emission lines may be those of the hot intracluster medium
surrounding the filament.

\section{Far Ultraviolet Emission Diagnostics}
The far ultraviolet waveband is a crucial region for exploring the heating
mechanism in these filaments. Recently, observations with the Hubble Space
Telescope have revealed far ultraviolet emission that shows a good spatial
correlation with emission-line nebulae (\citealt{Odeaetal04};
\citealt{Odeaetal10,Sparksetal09,Oonketal11}). Although this does indicate some
star formation is present in some regions, this is probably not relevant to the bulk of
the remote emission-line
filaments; in particular, we note that although hot stars are very efficient
at producing optical emission-line nebulae they are
unable to reproduce the crucial emission-line diagnostics seen in central
cluster galaxy filaments and referred to at the end
of Section 1. Recently, \citet{Sparksetal09} detected far ultraviolet emission
in the filamentary system around M87 using broad-band imaging. Follow-up spectroscopy
by \citet{Sparksetal12} has revealed this to be due to emission lines of
CIV$\lambda$1550 and HeII$\lambda$1640. They interpret these lines as being formed in
$10^5$K gas arising from thermal conduction between the hot $10^7$K and
cold $10^4$K phases.

\subsection{The continuum}
A key prediction of our particle heating model is the continuum that is expected
to be associated with the filaments. Since there is no incident photon continuum
(apart from the weak metagalactic background flux) the only continuum emission
associated with particle-heated filaments will be the diffuse continuum
generated by atomic processes within the filament. The Hydrogen two-photon
continuum is relevant to the far ultraviolet part of the spectrum.

\begin{figure}
\includegraphics[width=\columnwidth]{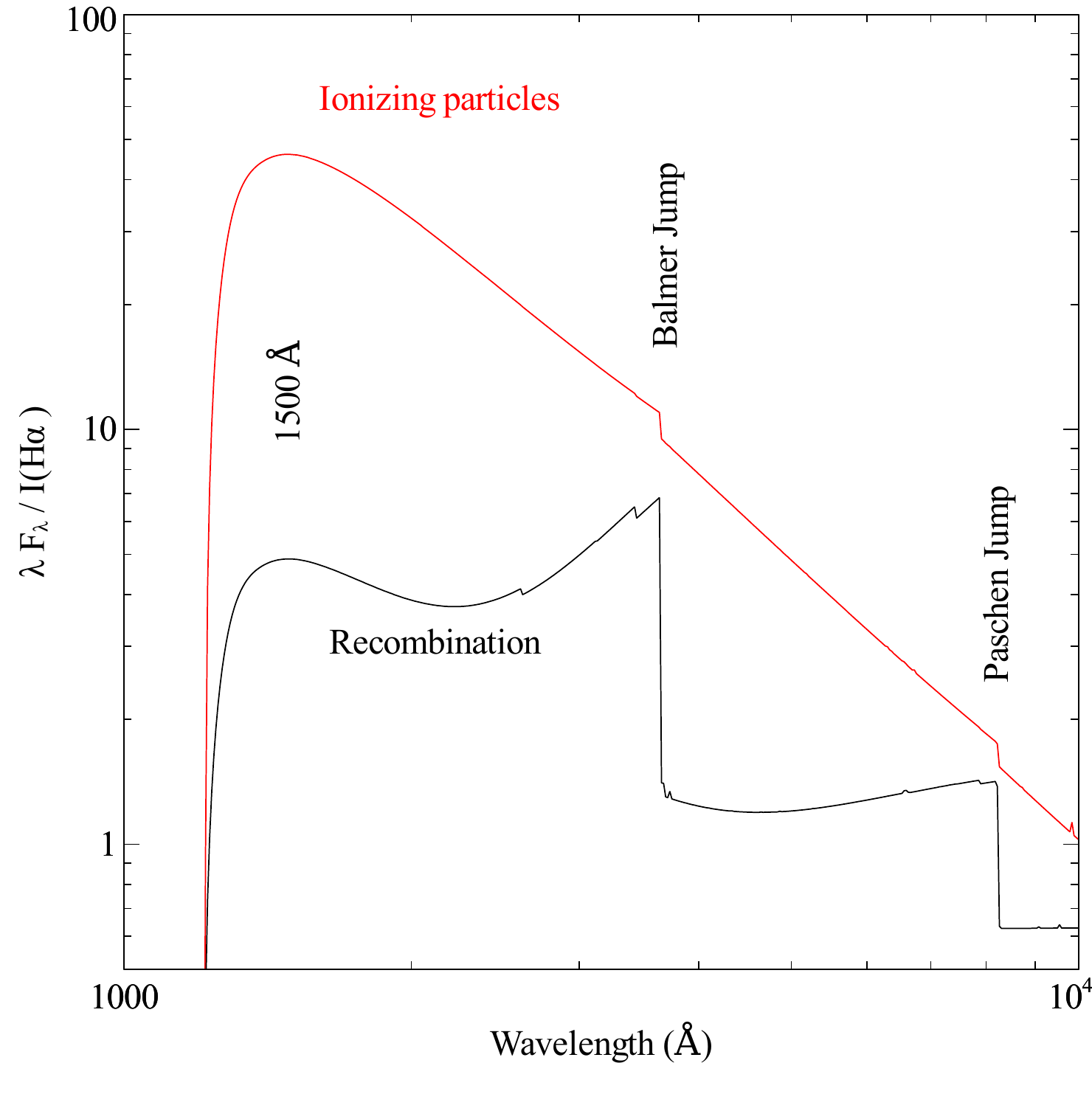}
\caption{Comparison between emission produced by a photoionized gas (black line)
and one energized by ionizing particles (red line).  The continuous emission is
given in terms of $\lambda F_{\lambda}$ (with units $\ergpccmps$) divided by the
intensity of H$\alpha$ in the same units.  As a result the vertical axis is
dimensionless, but could be changed to F$_{\lambda}$ / I(H$\alpha$) by dividing
by the wavelength at each point.  The ionizing particle case produces a far
stronger two-photon continuum due to excitation to the $n=2$ configuration by
secondary non-thermal particles. In this case, the two-photon continuum dwarfs
the broad recombination edges which are a hallmark of a photoionized gas and
could easily mimic the continuum of a hot star.}
\label{2nu}
\end{figure}

Tests of our particle heating model show that the region with a temperature near
T$\sim 10^4$ K dominates
the total H~I two-photon emission due to the high emissivity at that
temperature.  Fig.~\ref{2nu} shows the 1000-10000~\AA\ spectrum from a single
grid point in the particle heating flux / density plane with temperature $\sim
10^4$K in dimensionless units, $\lambda$F$_\lambda$ normalized by the H$\alpha$
flux from the same grid point.

The continuum spectrum produced by a low-density (n$_{\rm H}$ = 100 cm$^{-3}$)
photoionized gas is also shown in the same panel for reference. The physics of
line and continuum emission produced by photoionization followed by
recombination is covered in \citet{OsterbrockFerland06}. Case B conditions,
where the Lyman lines are optically thick and higher series such as Balmer and
Paschen are optically thin, should apply to the filaments.  In Case B roughly
one H$\alpha$ photon is produced for each H$^+$ recombination, and a third of
recombinations eventually populate the 2$s$ level that produces the two-photon
continuum.  The two-photon emissivity peaks, in terms of the photon flux at half
the energy of Ly$\alpha$, or $\sim 2432$ \AA, while in terms of energy flux the
emissivity peaks at a slightly shorter wavelength near $\lambda1500$\AA. 
Fig.~\ref{2nu} also shows two prominent recombination edges, the Balmer jump at $\sim
3646$\AA\ and the Paschen jump at $8204$\AA, which are a hallmark of a
photoionized 
gas.

A far stronger two-photon continuum is produced in the ionizing particle case
relative to the emitted H$\alpha$ flux.  Here the H~I emission is produced by
non-thermal secondary electrons that excite atomic hydrogen from the ground
configuration.  The largest cross-sections are for excitation to the $n=2$
configuration, which then produces Ly$\alpha$ and the two-photon continuum.  The
cross-sections for exciting to $n \ge 3$, which can decay and produce H$\alpha$,
are much smaller.  The result is that the two-photon continuum is roughly a
factor of 10 stronger, relative to H$\alpha$, than is produced by
photoionization followed by recombination.  The emission is so strong in the
ionizing particle case that the two-photon continuum dwarfs the recombination
Balmer and Paschen edges.  The blue / far ultraviolet spectrum is dominated by a
strong continuum that might be mistaken for emission from a hot star if the
cutoff at 1216\AA\ was not covered in a spectrum.

\subsection {The emission line of C~I $\lambda 1656$\AA}
\begin{figure}
\includegraphics[width=\columnwidth]{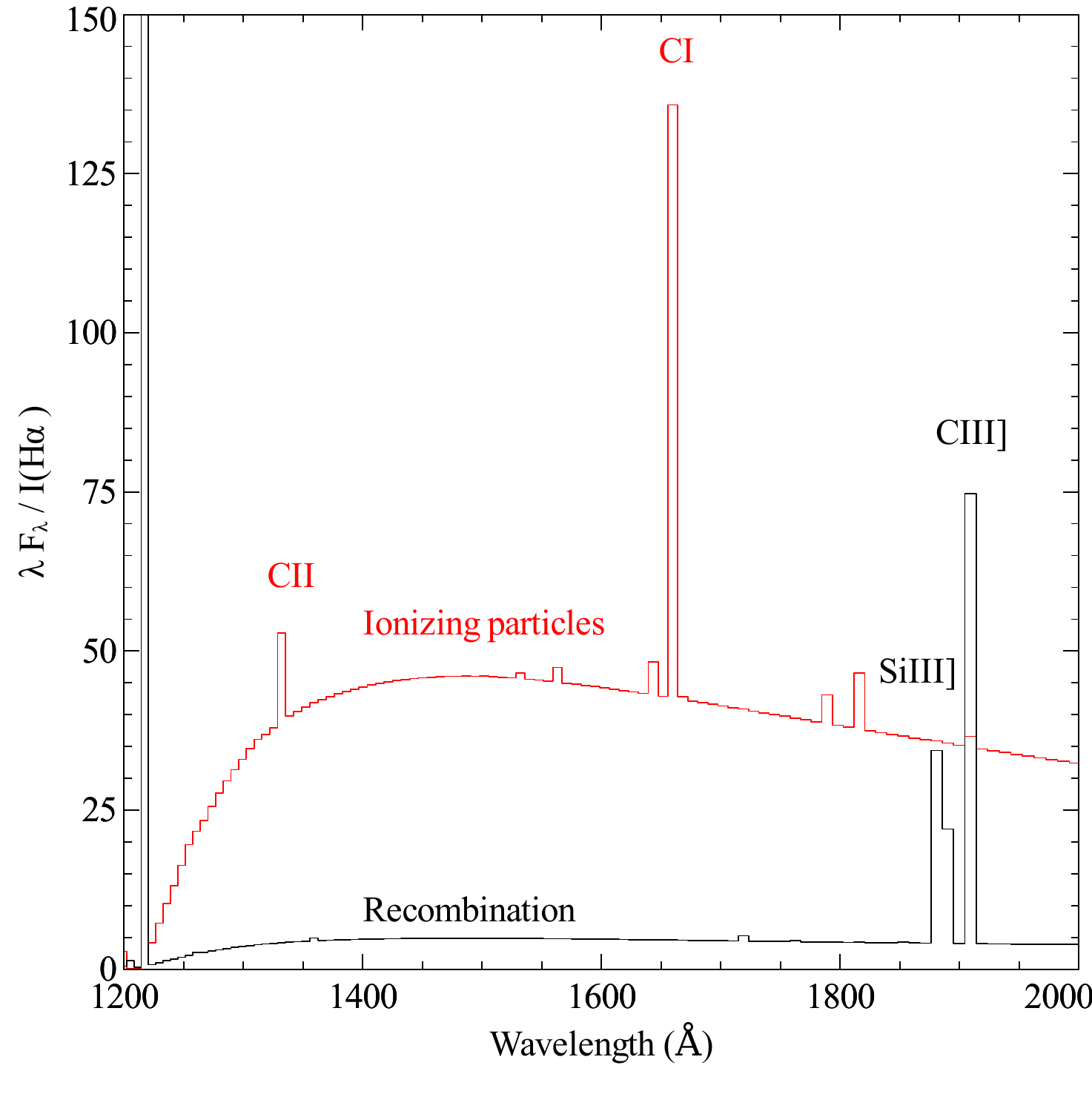}
\caption{Far ultraviolet region of the spectra presented in Fig.~\ref{2nu}. Note
that the contrast of the emission lines against the continuum is dependent on
the resolution of the spectrum and the intrinsic line width. Therefore the
strength of the emission lines, whilst correct in a relative sense between
emission lines in a spectrum is not to be simply read off from this spectrum.}
\label{ci}
\end{figure}

In Fig.~\ref{ci} we show the far ultraviolet part of the spectrum in more
detail, on a linear scale appropriate for comparison with observations, and
including line emission. (Note that the contrast of the emission lines against
the continuum is dependent on the resolution of the spectrum and the intrinsic
line width. Therefore the strength of the emission lines, whilst correct in a
relative sense between emission lines in a spectrum is not to be simply read off
from this spectrum.) The particle heating model gives rise to a strong emission
line from C~I at $\lambda1656$\AA\ that is not seen in the recombination
spectrum.

\citet{Dixonetal96} presented a far ultraviolet HUT spectrum of NGC~1275 taken
through a 9x116 arcsec slit passing through the galaxy nucleus at position angle
105$^\circ$. C~I $\lambda 1656$\AA\ at the redshift of NGC~1275 ($z=0.0176$)
agrees well with the wavelength of the feature marked ``Unk" in their figure 2
and listed as ``Unidentified" in their table 2. The signal-to-noise ratio for
this feature is low, and the slit is very large and passes through the bright
nucleus. Nonetheless these data are suggestive of a detection of C~I which could
be confirmed in future HST observations.

The reason that the C~I line is so strong in the particle heating model is due
to the presence of suprathermal electrons in the regions where Carbon is
neutral. Such suprathermal electrons are needed to collisionally excite this
line which arises from an energy level ~7.5~eV above the ground state and which
would not normally be populated by a thermal distribution of particles where the
temperature is low enough for Carbon to be neutral. This line and its formation
mechanism is analagous to the [NI] doublet seen in the optical waveband at
$\lambda5199$\AA.

We note that the recombination spectrum shows lines of semi-forbidden
SiIII]$\lambda 1892$ and CIII]$\lambda1909$ which the particle heating model
does not.

\section{Observations and Data Reduction}
\label{obsred}
In order to confront the particle heating model predictions for the ultraviolet
continuum with observations we searched the Multimission Archive at the Space
Telescope Science Institute (MAST) and Hubble Legacy Archives for ultraviolet
spectra of Region 11 or images covering the region of the Horseshoe in NGC~1275.
No ultraviolet spectra of this region are available, but there do exist ACS
Solar Blind Channel images. We selected an image (data association id JA2703010)
taken with the F140LP filter as most appropriate for this study due to its
inclusion of the 1500\AA\ peak of the Hydrogen two-photon continuum.
Fig.~\ref{f140lpfilt} shows the relative system thoughput for this configuration,
with the spectrum predicted by the particle heating model presented in
Fig.~\ref{ci} overlaid.

\begin{figure}
\includegraphics[width=\columnwidth]{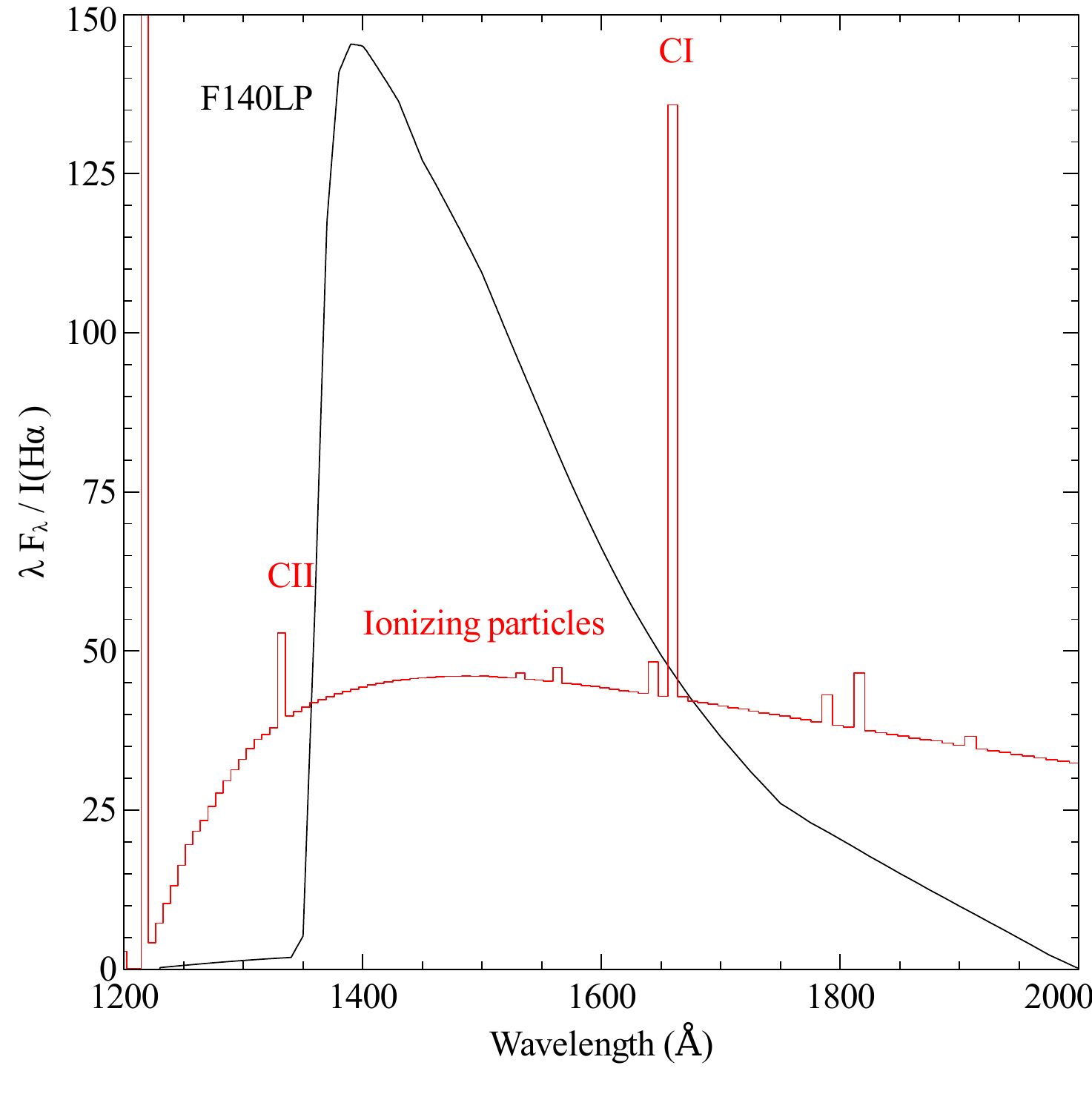}
\caption{Relative system thoughput for ACS/SBC with F140LP filter (solid black line),
calculated from the
{\sc stsdas} task {\sc bandpar}. Overlaid in red, as a histogram plot,
is the spectrum predicted by the particle heating model shown in Fig.~\ref{ci}.
}
\label{f140lpfilt}
\end{figure}

Other supporting images used in this paper include our own ACS/WFC data (F625W)
(to estimate the H$\alpha$ flux) already published in \citet{Fabianetal08} and
further WFPC2/F702W (to confirm the H$\alpha$ flux) and WFPC2/F814W (galaxy
continuum) images downloaded from MAST. Details of the observations used are
listed in Table~\ref{obslog}.

The reduction of the F625W data have already been discussed in
\citet{Fabianetal08}. The data downloaded from MAST is already fully reduced and
we have used the final drizzled data products in this work. Data taken with HST
at different epochs can have small offsets in the world coordinate systems due
to the different guide stars used by the telescope. To align the zero points of
the world coordinate systems of the data downloaded from the archive with our
previously published images we applied a linear shift to register point sources
with the ACS/F625W frame using the program {\sc gaia} from the Starlink Software
Collection.

Fig.~\ref{dataimages} shows a set of three images of an area of sky in the
region of the Horseshoe filament. From top to bottom these images were taken
through the F140LP, F625W and F814W filters. The two regions that we have
analysed in this paper are indicated: the upper left circle (Region A) is
centred on the bright knot of H$\alpha$ emission at the base of the Horseshoe
and corresponds closely to Region 11 in the work of \citet{Conseliceetal01},
while the lower right circle is Region B that we use to assess the background
count rates.

\begin{table*}
\caption[]{Log of Observations. $^*$ Due to the mosaicing of several pointing,
not all of the image has the same exposure time. This is the exposure time for
the Region A area.}
\begin{tabular}{cccc}
\hline
Instrument&Data&Origin&Exposure\\
&&&Seconds\\
\hline
ACS/SBC/F140LP&ja2703010\_drz\_sci.fits&MAST/Hubble&2552\\
ACS/WFC/F625W&N/A&Fabian et al 2008&10751$^*$\\
WFPC2/F702W&hst\_06228\_04\_wfpc2\_f702w\_wf\_drz.fits&MAST/Hubble Legacy&4700\\
WFPC2/F814W&hst\_11207\_a1\_wfpc2\_f814w\_wf\_drz.fits&MAST/Hubble Legacy&1600\\
\hline
\end{tabular}
\label{obslog}
\end{table*}

\begin{figure}
%
\includegraphics[width=\columnwidth]{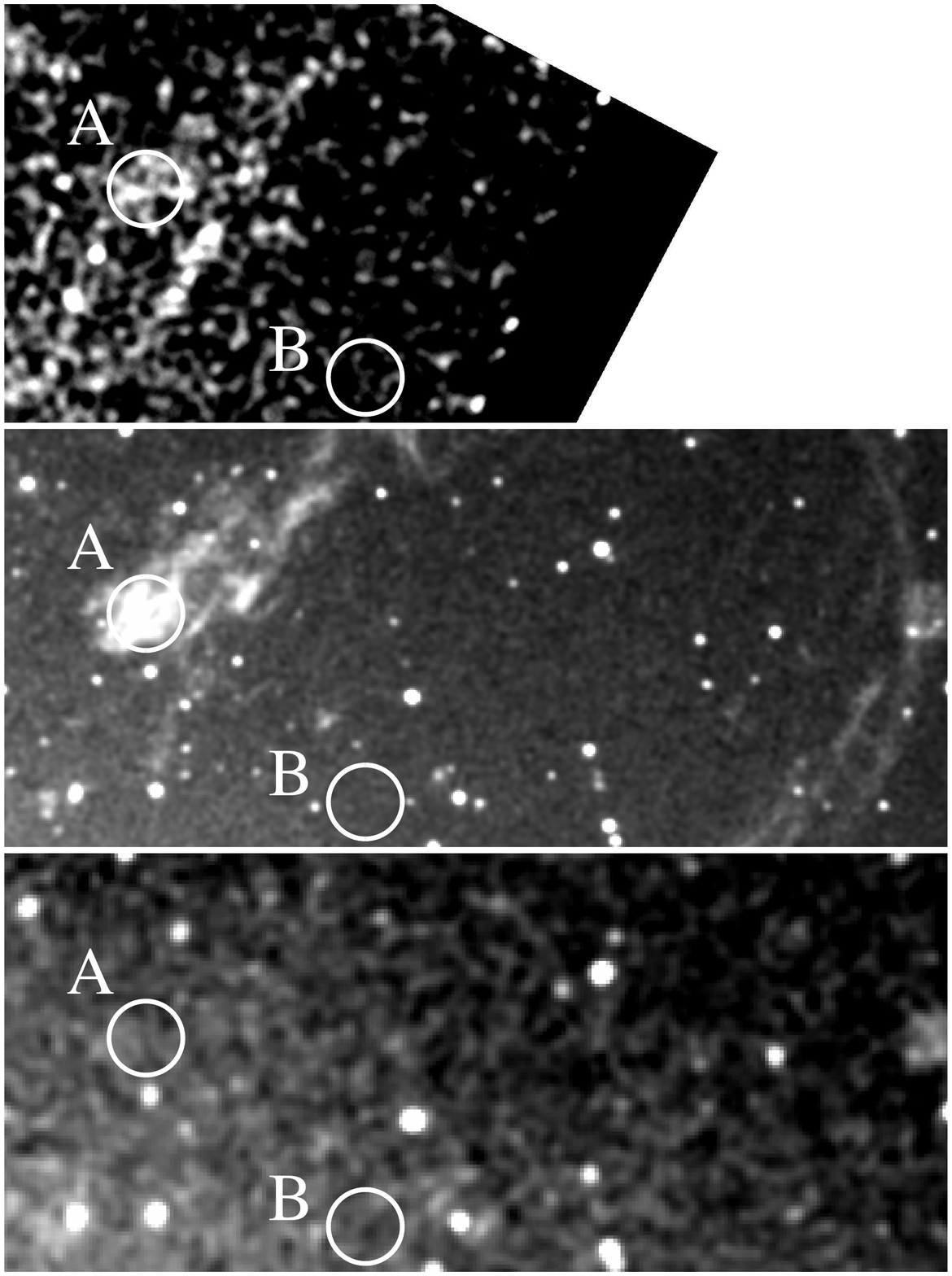}
\caption{HST images of the north-west part of the NGC~1275 emission-line
nebula, centred on the Horseshoe filament near RA(2000)=03:19:44.51
Dec(2000)=+41:31:32.5.
Each frame is 25 arcsec $\times$
11 arcsec; North is up, East to the left. From top to bottom the images are
taken though filters F140LP, F625W
and F814W. The images are displayed on a linear scale and have been smoothed
with a gaussian kernel of sigma
0.275, 0.15 , 0.3 arcsec respectively. The circles show regions A and B referred
to in the text.}
\label{dataimages}
\end{figure}

\section{Analysis}
Examination of the images presented in Fig.~\ref{dataimages} reveals that the
brightest part of the filamentary structure at the base of the Horseshoe seen in
the F625W (emission line) image also shows up in the F140LP (ultraviolet) image.

In the following analysis we use a 1 arcsec radius circular aperture centred on
the bright knot of emission located at
RA(2000)=03:19:45.203, Dec(2000)=+41:31:33.41 as our best estimate of Region 11
of \citet{Conseliceetal01}.

\subsection{Background subtraction}
The filamentary structures seen in the nebula around NGC~1275 are superimposed
on the underlying smooth emission from
the galaxy stellar population. We note that in some spectral bands, in some
spatial regions, emission and absorption from the infalling
8000~\kmps\ system (\citealt{Minkowski57,BurbidgeBurbidge65,Rubinetal77})
provides additional complexity. The Horseshoe region, however, appears to be
clear of any contamination from the infalling system.

The F140LP data has not had any background subtracted from it, whereas the F625W
data which is part of the mosaic published by \citet{Fabianetal08} has had an
approximate background, estimated from regions close to the edge of the
component frames, subtracted.

In order to analyse the emission from Region 11 we need to remove not only the
sky background, but also the underlying galaxy continuum. Under the assumption
that the profile of the galaxy, and the flatness of the sky emission is the same
in different filters, we can use an image free from emission lines to define a
background Region B that has the same galaxy plus sky count rate as the region
of interest, A. The F814W image fulfils this criterion well. The emission in the
F814W filter which has a passband from approximately 7000\AA\ to approximately
9600\AA\ is expected to be dominated by light from the underlying elliptical
galaxy due to the filter's very red bandpass and the fact that there are no
strong emission lines at the redshift of NGC~1275 that occur within its
bandpass. We confirmed that no filamentary structure is present in this filter
under smoothing using a range of kernel sizes. A circular region, B, of radius 1
arcsec, located at RA=03:19:44.674, Dec=+41:31:28.31, within 10 arcsec of
Region 
A, was found to have a count rate very similar to that of Region A (see Table
\ref{ctrts}) and is used as our background region in the other filters. We note
that this method of background subtraction also removes any diffuse component of
H$\alpha$ emission that might be present in the F625W image.

In Table~\ref{ctrts} we show the count rates measured in regions A, B  as well
as the net count rate A-B in the four filters.

\begin{table}
\caption[]{Measured Count Rates. $^*$ SBC data are in units of counts/s.}
\begin{tabular}{lrrr}
\hline
Instrument&Region A&Region B&Region A-B\\
&e$^-$/s&e$^-$/s&e$^-$/s\\
\hline
ACS/SBC/F140LP$^*$&0.3667&0.2427&0.12\\
ACS/WFC/F625W&47.463&10.438&37.03\\
WFPC2/F702W&18.52&8.19&10.33\\
WFPC2/F814W&57.47&57.24&0.23\\
\hline
\end{tabular}
\label{ctrts}
\end{table}

The F140LP data shows a background subtracted count rate of 0.12 e$^{-}$/s in
Region A, and yields approximately an 8$\sigma$ detection. The F625W shows a net
count rate of 37.0 e$^{-}$/s. These counts come though a broad band filter, and,
on their own are not simple to interpret.

\subsection{Extinction corrections}
NGC~1275 lies at a Galactic Latitude of $-13^{\circ}$ and is therefore
relatively highly reddened by dust in our Galaxy. Many values of extinction have
appeared in the literature. The NASA Extragalactic Database (NED) lists
E(B-V)=0.163; \citet{Conseliceetal01} use E(B-V)=0.17. We adopt this latter
value in the current work and use the D Welch's web form at
\url{http://dogwood.physics.mcmaster.ca/Acurve.html} to calculate the absorption
at specific wavelengths. This calculator, which is based on the formulae in
\citet{Cardellietal89} requires, as input, the values of R$_{\rm V}$, the ratio
of total to selective absorption at V and A$_{\rm V}$ the total absorption in
magnitudes at V. We adopt the commonly used value of R$_{\rm V}$=3.1, and
A$_{\rm V}$=0.527, appropriate to E(B-V)=0.17.

\subsection{Ultraviolet continuum flux}
The ACS exposure time calculator
(\url{http://etc.stsci.edu/etc/input/acs/imaging/}) with the two-photon
continuum shape produced from {\sc cloudy}, (see Section~6 and
Fig.~\ref{contspec}) redshifted and absorbed by a Galactic extinction of
E(B-V)=0.17, indicates that for a uniform surface brightness source filling a 1
arcsec radius aperture, the observed count rate of 0.12 ct/s for Region A in the
F140LP band is produced by a source with an intrinsic (extinction corrected)
flux density of $1.28\times10^{-17}\ergpscmpspa$ at 1500\AA.

We note that the morphology of the emission within Region A is close to a
uniform surface brightness source, but not completely filling the aperture and
with some bright knots. Running the exposure time calculator again, but assuming
a point source morphology for Region A, yields a flux of
$1.40\times10^{-17}\ergpscmpspa$ for the same count rate, ie a flux that is 8
per cent brighter. Since the actual morphology is between the two extremes of a
point source and a constant surface brightness source, but more like a constant
surface brightness source than a point source we adopt the count rate to flux
conversion factor for a uniform surface brightness source and note that the
emission line flux would be brighter, but by a lot less than 8 per cent if the
actual morphology were used.

\subsection {H$\alpha$ line flux}
The ACS exposure time calculator
(\url{http://etc.stsci.edu/etc/input/acs/imaging/}) indicates that for a single
emission line at 6678.5\AA\, the redshifted wavelength of H$\alpha$ the observed
count rate of 37.03 e$^-$/s is produced with a line flux of
$5.48\times10^{-15}\ergpscmps$. After correcting by a factor of 1.474 to account
for Galactic reddening this gives an intrinsic flux of
$8.1\times10^{-15}\ergpscmps$. If a point source had been assumed the observed
flux would be $5.71\times10^{-15}\ergpscmps$, an increase of 4 per cent.

The F625W filter admits flux from the [OI], [NII] and [SII] doublets as well as
the H$\alpha$ line (continuum emission in the filter band has been subtracted).
In order to correct for these additional emission lines we use the relative
strengths of these lines with respect to H$\alpha$, published by
\cite{Ferlandetal09}. All the lines lie at wavelengths where the thoughput of
the F625W filter is near its maximum. The flux in the H$\alpha$ line is found to
be 1/2.6 or 0.38 times the total line flux. The implied H$\alpha$ line flux is
therefore $3.1\times10^{-15}\ergpscmps$.

\subsubsection{Comparison with the emission line flux in
\protect{\citet{Conseliceetal01}}}
We note here that Region 11 in the work of \citet{Conseliceetal01} is almost
identical to our Region A. Their table~1 lists a flux for H$\alpha$ of
$2.25\times10^{-14}\ergpscmps$. We assume that this flux includes the
[NII]$\lambda\lambda6548,6584$ lines since they lie within the bandpass of their
filter and there is no mention in the paper of correcting for the contribution
from the [NII] lines. We further assume that the listed flux has been corrected
for Galactic extinction following the prescription in their Section 2.

The passband listing for the KPNO filter KP1495 available at
\url{ftp://ftp.noao.edu/kpno/filters/4inplots/kp1495.eps} used by
\citet{Conseliceetal01} shows that the redshifted H$\alpha$ line lies very close to the
peak of the filter transmission. The [NII]$\lambda\lambda6548,6584$ lines also lie within
7 and 3 per cent of the maximum transmission respectively. We therefore assume that
all the counts come from the wavelength of H$\alpha$ and note that the
[NII]$\lambda6548$ carries only one third of the flux in [NII]$\lambda6584$ so the
error in its calibration will be very minor.

Given the line ratios listed above, the extinction corrected H$\alpha$ flux in
\citet{Conseliceetal01} is inferred to be 1/1.95 times the total flux or
$1.2\times10^{-14}\ergpscmps$. This is a factor of 3.9 larger than our
extinction corrected flux.

In order to determine whether there is a flux calibration problem with our F625W
data we identified four stars from the Hubble Guide Star Catalog (GSC) (version
2.3) which were faint enough to not be badly saturated, but bright enough to
have magnitudes measured in several bands. Spectral energy distributions from a
library of stellar models were convolved with the GSC bandpasses and fitted to
the (j-N) colours of these stars for us by P Hewett. The best fitting spectral
energy distribution, normalized to the F (red) magnitude was then input to the
ACS imaging exposure time calculator
(\url{http://etc.stsci.edu/etc/input/acs/imaging/}) to predict the count rate
expected in the F625W data. Table~\ref{stdstars} lists the stars together with
their magnitudes and their predicted and observed count rates (both in circular
apertures of radius 0.5 arcsec). In all cases the agreement is better than 33
per cent, (which is well within the uncertainties of the catalogue magnitudes)
indicating that the factor of four 
discrepancy between our fluxes and those published by \citet{Conseliceetal01} is
likely not due to the F625W data. We note here that these measurements were made
on intermediate product {\emph single\_sci} frames rather than the final
mosaiced image since some of the stars are bright enough that the cosmic ray
rejection algorithms applied to the final mosaic clip up to 15 per cent of the
counts in the brighter stars. We spot checked the count rate in Region A in the
mosaic and in the single\_sci frames and found agreement within 5 per cent.

\begin{table*}
\caption[]{Guide Star Catalogue Stars Data. Col 1: Catalogue star name; cols
2-5: magnitudes and uncertainties in the j, V, F and N bands; cols 6-7:
predicted and observed count rates in circular aperture with radius 0.5 arcsec;
col 8: percentage error between predicted and observed count rates.}
\begin{tabular}{cccccccc}
\hline
Name&m$_{\rm j}$&m$_{\rm V}$&m$_{\rm F}$ & m$_{\rm N}$
&F625W(pred)&F625W(obs)&100$\times$(pred-obs)/pred\\
&&&&&e$^-$/s&e$^-$/s&per cent\\
\hline
NCJ1022694&19.53$\pm$0.44&19.13$\pm$0.65&18.88$\pm$0.46&18.85$\pm$0.55&425.1&400
.1&-6\\
NCJ1022527&19.83$\pm$0.45&19.27$\pm$0.77&19.10$\pm$0.47&19.11$\pm$0.61&334.6&222
.9&-33\\
NCJ1022851&20.77$\pm$0.48&19.11$\pm$0.63&19.02$\pm$0.47&18.83$\pm$0.56&315.0&232
.7&-26\\
NCJ1022049&19.71$\pm$0.45&19.19$\pm$0.73&18.95$\pm$0.46&18.89$\pm$0.56&389.4&324
.9&-17\\
\hline
\end{tabular}
\label{stdstars}
\end{table*}

As an extra check on our calibration we have made a separate measurement of the
emission-line flux from a WFPC2 data set taken with the F702W filter which
covers the Horseshoe region. This filter also admits the [OI], [NII], and [SII]
doublets as well as the H$\alpha$ line, so should yield a background subtracted
flux very similar to th ACS F625W data. We note that these data were downloaded
from the Hubble Legacy Archive and that the WFPC2 data products in this archive
have been processed through the drizzling software using non-standard
parameters, yielding images in units of electrons per second, rather than the
default DN per second (\url{http://hla.stsci.edu/hla_faq.html}). We note also
that the Fits header keyword BUNIT remains set to COUNTS/S despite the data
being in units of electrons/s. Using the WFPC2 exposure time calculator at
\url{http://www.stsci.edu/hst/wfpc2/software/wfpc2-etc-extended-source-v40.html}
the net count rate of 10.33 cts/s for Region A in this instrument yields a line
flux of $5.
14\times10^{-15}\ergpscmps$ in close agreement ($\sim 7$ per cent) with the ACS
F625W data for the same region ($5.48\times10^{-15}\ergpscmps$).

\citet{Privonetal08} noted that extended emission-line regions in radio galaxies
had lower fluxes when observerved with HST WFPC2 compared with ground-based
observations. They attributed this to a higher surface brightness detection
limit due to the smaller pixels in the HST data. We note here that throughout
this work we are measuring fluxes in the same
size aperture as the fluxes presented by \citet{Conseliceetal01} and therefore
expect to obtain a consistent flux despite the lower surface brightness
sensitivity of the HST data due to the smaller pixel size.

\section{Continuum Emission from the Particle heating Model}
\begin{figure}
\centerline{\includegraphics[width=0.47\columnwidth]{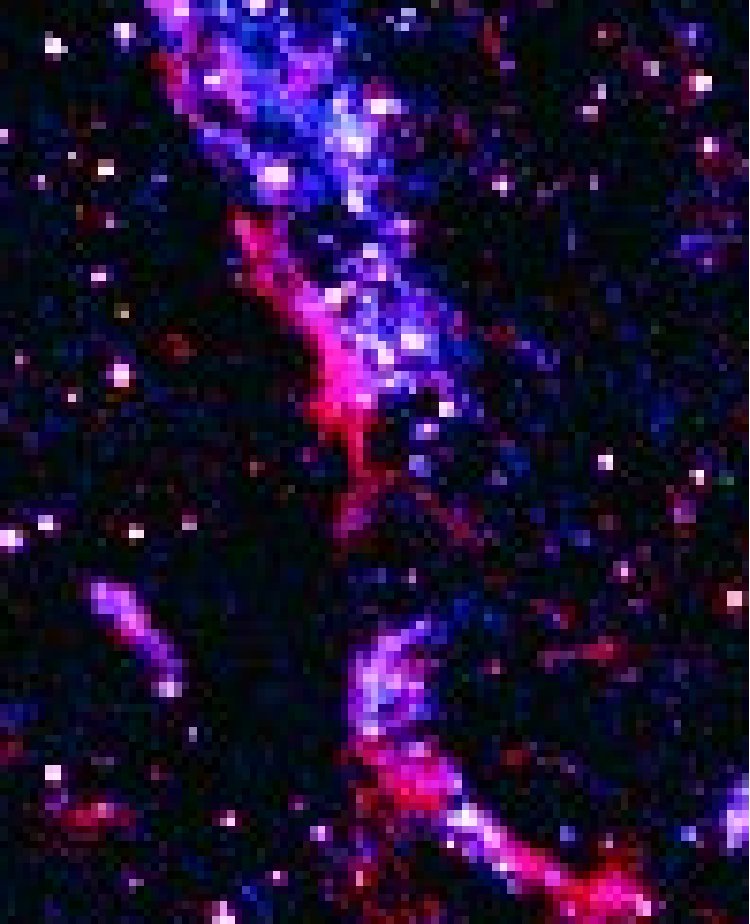}
\includegraphics[width=0.5\columnwidth]{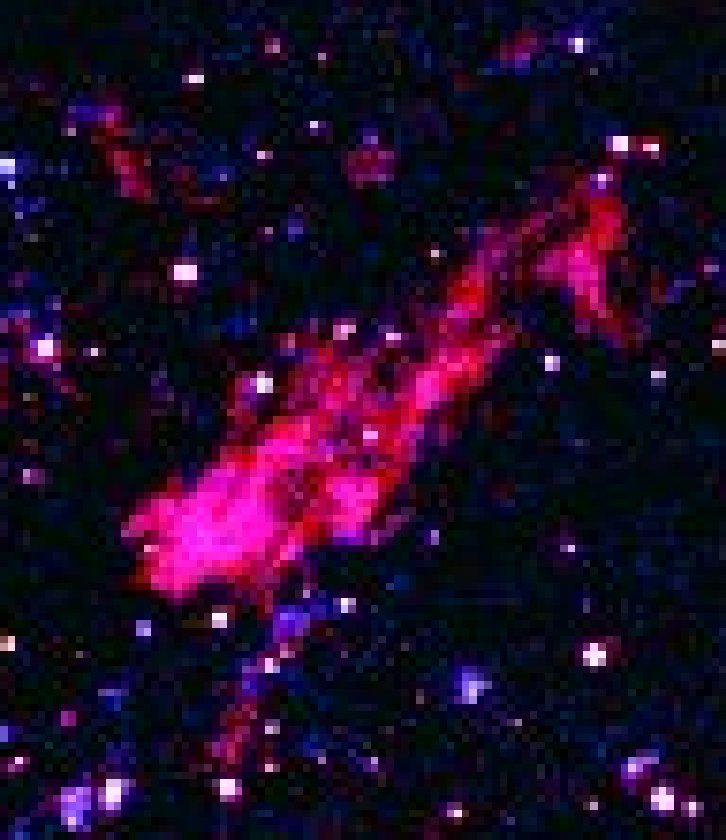}}

\caption{Comparison of two regions in the NGC~1275 nebula from observations
published by \citet{Fabianetal08}. The red channel (F625W filter) includes
emission from lines of [OI], H$\alpha$, [NII] and [SII]; the blue channel (F435W
filter) shows mostly blue continuum although it does include [OII]$\lambda3727$
emission at a point in the filter transmission that has 40 per cent of the peak
transmission. The left panel shows part of the South-West filament where there
are nearby hot stars while the right panels shows the Horseshoe filament knot
region with very few hot stars.}
\label{HSTregioncompare}
\end{figure}

In this section we consider the interpretation of the continuum emission in
terms of the particle heating model presented by \citet{Ferlandetal09}. Hot
young stars are unable to reproduce the optical emission-line ratios as
discussed in Section \ref{intro}. We also show here in
Fig.~\ref{HSTregioncompare} a comparison of two regions of the NGC~1275 nebula
from the observations published by \citet{Fabianetal08}. These images are colour
coded such that H$\alpha$ emission is red and blue continuum (F435W filter) is
blue. Although there are some regions in the nebula where a luminous blue
continuum is seen close to the H$\alpha$ emission (left panel, South-West
filament) the Horseshoe filament knot (right panel) does not show such luminous
blue emission, further suggesting that hot stars are not involved in exciting
the emission-line gas.

\begin{figure}
\includegraphics[width=\columnwidth]{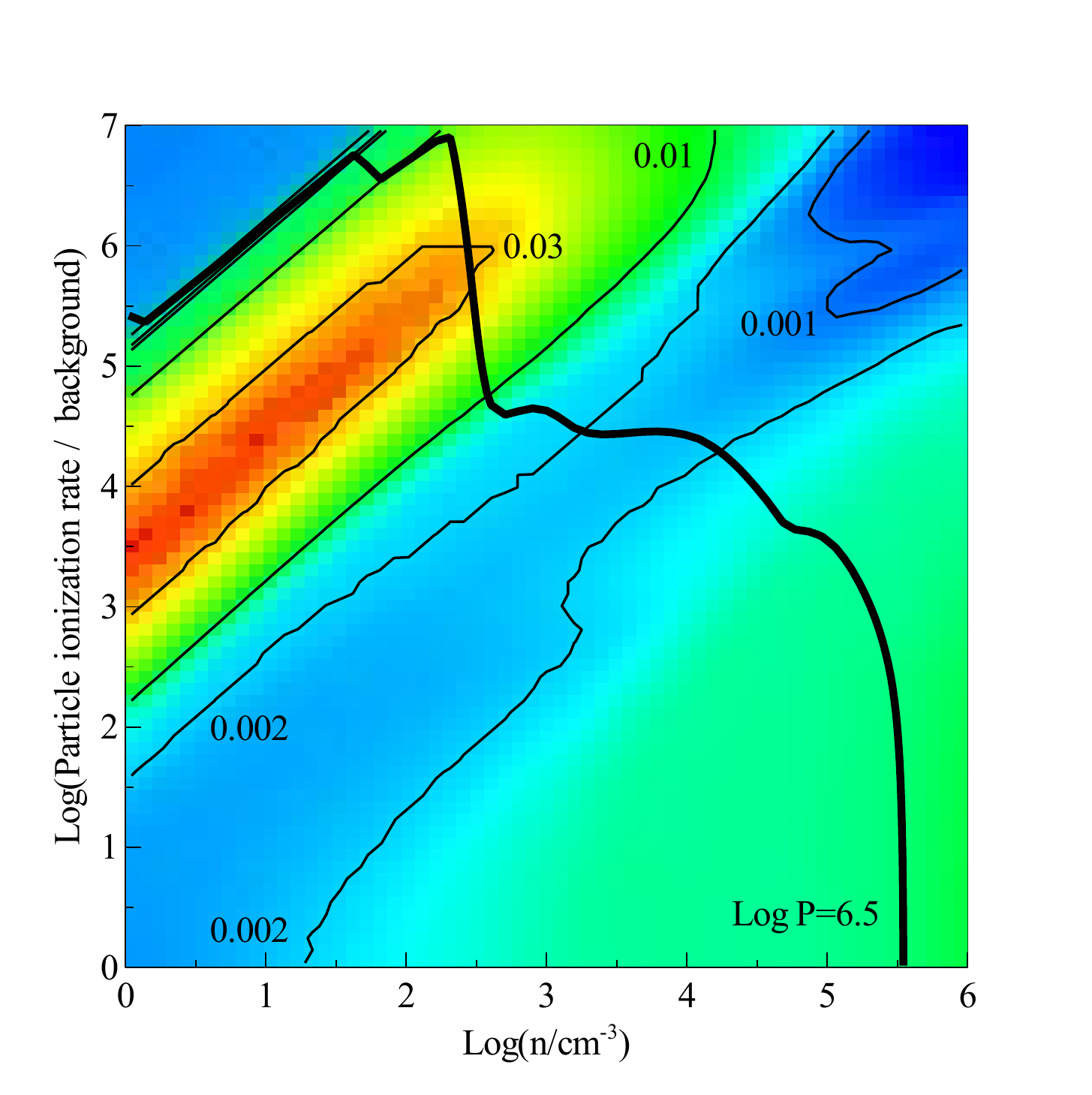}
\caption{R = F$_{1500}$ / I(H$\alpha$) ratio for particle heating models
explored by \citet{Ferlandetal09}. Contour values are indicated. The locus of
points with pressure P=$10^{6.5}\pccmK$ is shown by the thick black line.}
\label{twophotrat}
\end{figure}

In the particle heating models of \citet{Ferlandetal09}, the relative amounts of
gas at different temperatures was set using a power-law to describe the
cumulative filling factor of gas at a range of temperatures, their equation 14.
The power-law slope ($\alpha$=-0.35) was set to match the ratio of the
H$_2$~12.28$\mu$m / H$\alpha$ in Region 11 of the NGC~1275 nebula, but the
H$\alpha$ flux used to set that ratio was derived from the
\citet{Conseliceetal01} data. The result from the previous section that suggests
the fluxes from those data are too high by a factor of 3.9 has consequences for
our predictions from this model. Reducing the H$\alpha$ flux by a factor of 3.9
requires a power-law slope $\alpha=+0.2$ to match the new value of
H$_2$~12.28$\mu$m / H$\alpha$=0.12. A full treatment of the changes to the
predictions from the particle heating model that this change of slope causes is
beyond the scope of this paper and will be treated in a future paper.

For this paper, we have re-run the particle heating models explored by
\citet{Ferlandetal09}, using the development version of {\sc cloudy} (newmole
branch, revision 5792), in order to save the predictions for the continuum
intensity as well as the H$\alpha$ flux, and use the new value of $\alpha=0.2$.

The brightest part of the continuum is at $\sim1500$\AA\ so we have predicted
the flux level expected in the Horseshoe knot at this wavelength. To set the
normalization of the continuum emission we consider the ratio of weighted
emissivities in the continuum at 1500\AA\ and the H$\alpha$ line.

\begin{figure}
\includegraphics[width=\columnwidth]{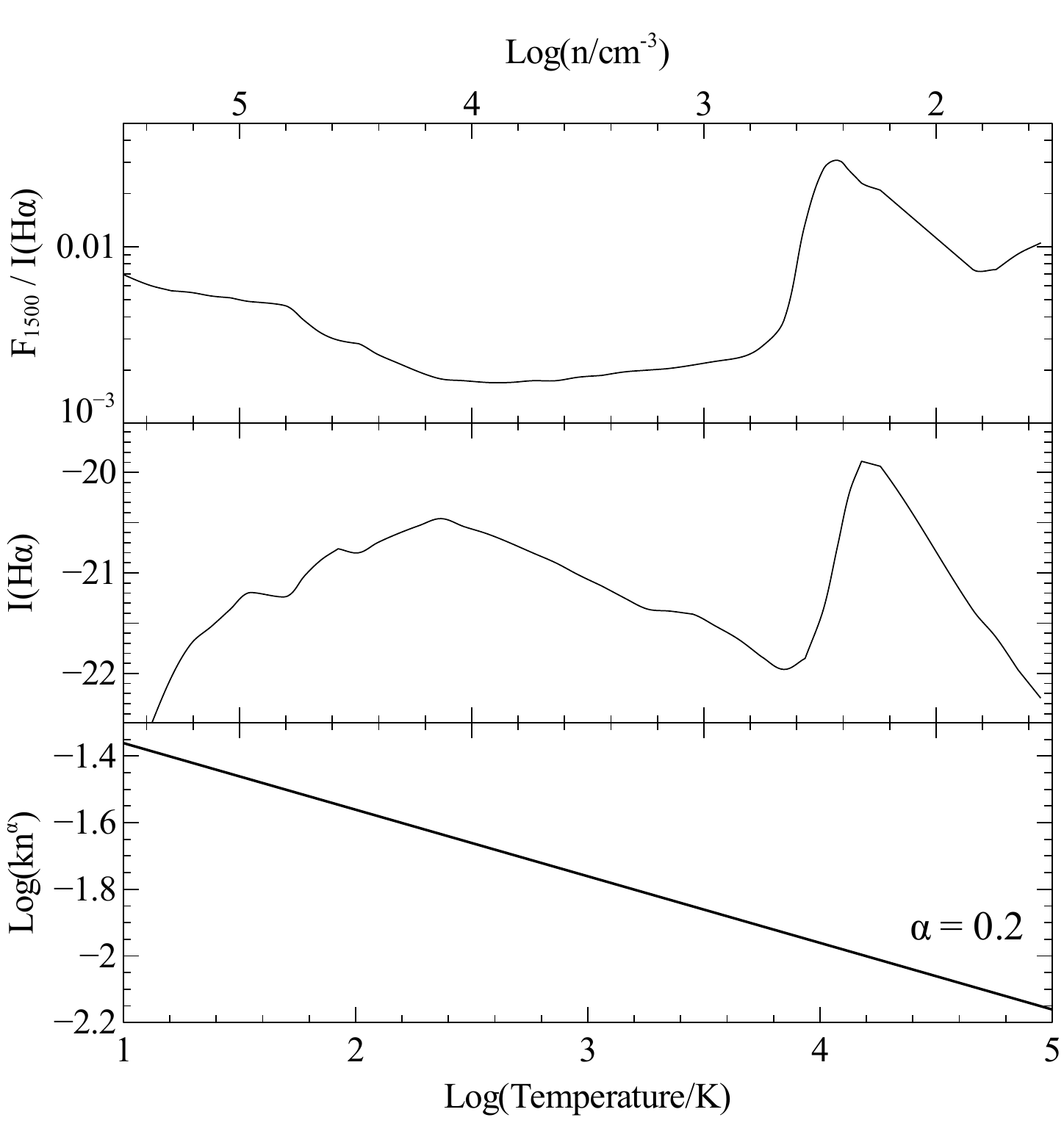}
\caption{Values of R = F$_{1500}$ / I(H$\alpha$) (top panel), emissivity of
H$\alpha$ (middle panel), and weighting function with $\alpha$=0.2 (lower panel)
as a function of temperature for the particle heating models explored by
\citet{Ferlandetal09}, for the locus of grid points with pressure
P=$10^{6.5}\pccmK$.}
\label{twophotrattemp}
\end{figure}

In Fig.~\ref{twophotrat} we show how the ratio R=F$_{1500}$/I(H$\alpha$) varies
over the entire particle flux / Hydrogen density plane. Values ranging between
$4\times10^{-4}$ and $3.2\times10^{-2}$ are produced in various parts of
parameter space. It should be noted however that the emissivity and weighting
factors vary considerably with temperature. The part of the diagram that we are
interested in is the locus of models for which the gas pressure is
P=$10^{6.5}\pccmK$, shown by the thick black line. Fig.~\ref{twophotrattemp}
shows how the ratio R, the emissivity of the H$\alpha$ line and the weighting
function with the new value of $\alpha$=0.2, vary as a function of temperature
along this constant pressure line.

To understand how the continuum to line ratio R varies across
Fig.~\ref{twophotrattemp} we note that the H$\alpha$ intensity is proportional
to the rate that the $n=3$ configuration is populated while the two-photon
continuum is proportional to the rate that the $2s$ level is populated
multiplied by the fraction of these populations that decay to $1s$ rather than
undergoing a collision to $2p$ (which then produces L$\alpha$). 

The $2s$ term can be populated by collisions from $1s$ (mainly due to
suprathermal electrons), recombination (either directly to $2s$ or indirectly by
capture to higher levels followed by cascade to $2s$), and by collisions from
$2p$.  The relative importance of these processes depends on the relative
abundances of suprathermal electrons, thermal electrons, and protons.

The variation in R is due to the remarkable changes in the physical conditions
as we go across Fig.~\ref{twophotrattemp}.  The gas is almost fully molecular in
the cold regions to the left, while it is moderately ionized in the warmest
regions to the right.  As an example, we consider three regions in
Fig.~\ref{twophotrattemp}, the
two regions of peak H$\alpha$ intensity and the lowest temperature point.   At
the highest temperature peak, near log T = 4.2, Hydrogen has a density of
$\sim200\pccm$ and is 22 per cent ionized with nearly all the rest atomic.  At
the log T = 2.4 peak the Hydrogen density is much higher, $10^4\pccm$ but the
gas is nearly entirely atomic.  The ionized fraction is only 0.2 per cent.  In
the highest density case, $n\sim3.2\times 10^{5}\pccm$ the gas is mostly
molecular (91 per cent) with a tiny ionized fraction ($\sim 10^{-5}$ per cent). 
The curious result of this is that as the total Hydrogen density increases the
electron density decreases, being $n_e =$ 48.4, 23.2, and 0.5 $\pccm$ at the
three 
points.  The resulting mix of physical processes causes the ratio R to vary in a
non-monotonic manner.

To calculate the ratio R from the model we make a weighted sum of the H$\alpha$
and F$_{1500}$ continuum emissivities along this constant pressure line between
log temperatures of 0.5 to 4.5. The weighting factors are set by the power-law
slope $\alpha$. At some temperatures the gas is thermally unstable
(\citealt{Ferlandetal09}) and must be excluded from the sum. These regions occur
between values of log temperature between 1.5 and 1.9, and between 3.5 and 3.8.
The final ratio, R is then determined by dividing the weighted emissivity in the
continuum F$_{1500}$ by the weighted emissivity of H$\alpha$. We obtain a value
of 0.0082.

Fig.~\ref{contspec} shows the predicted weighted spectral distribution of the
continuum emissivity from the particle heating model over the wavelengths
1000-5000\AA. There are several components to this spectrum. The brightest one
is due to the Hydrogen two-photon continuum which peaks in the ultraviolet at
$\sim1500$\AA\ and has a sharp cutoff at 1216\AA. Compared with Fig.~\ref{2nu}
there are other contributions which come from the free-bound emission of
Hydrogen, and Helium. The peaks at $\lambda\lambda2600,3433$\AA\ are due to HeI
edges while the feature at $\lambda3646$\AA\ is a blend of a HeI edge and the
Hydrogen Balmer edge. We note that these peaks are much sharper than in the case
of a recombination spectrum (Fig.~\ref{2nu}) because they originate in much
colder gas near the peak in H$\alpha$ emissivity at few $\times100$K
(Fig.~\ref{twophotrattemp}). Such sharp edges are a characteristic prediction of
our particle heating model. They are not normally observed in the spectra of
emission-line 
nebulae, but have been seen in the shell around nova DQ Her
(\citealt{Williamsetal78,Ferlandetal84}). For further discussion of the continua
see Section 3 of this paper and  \citet{OsterbrockFerland06}.

\begin{figure}
%
\includegraphics[width=\columnwidth]{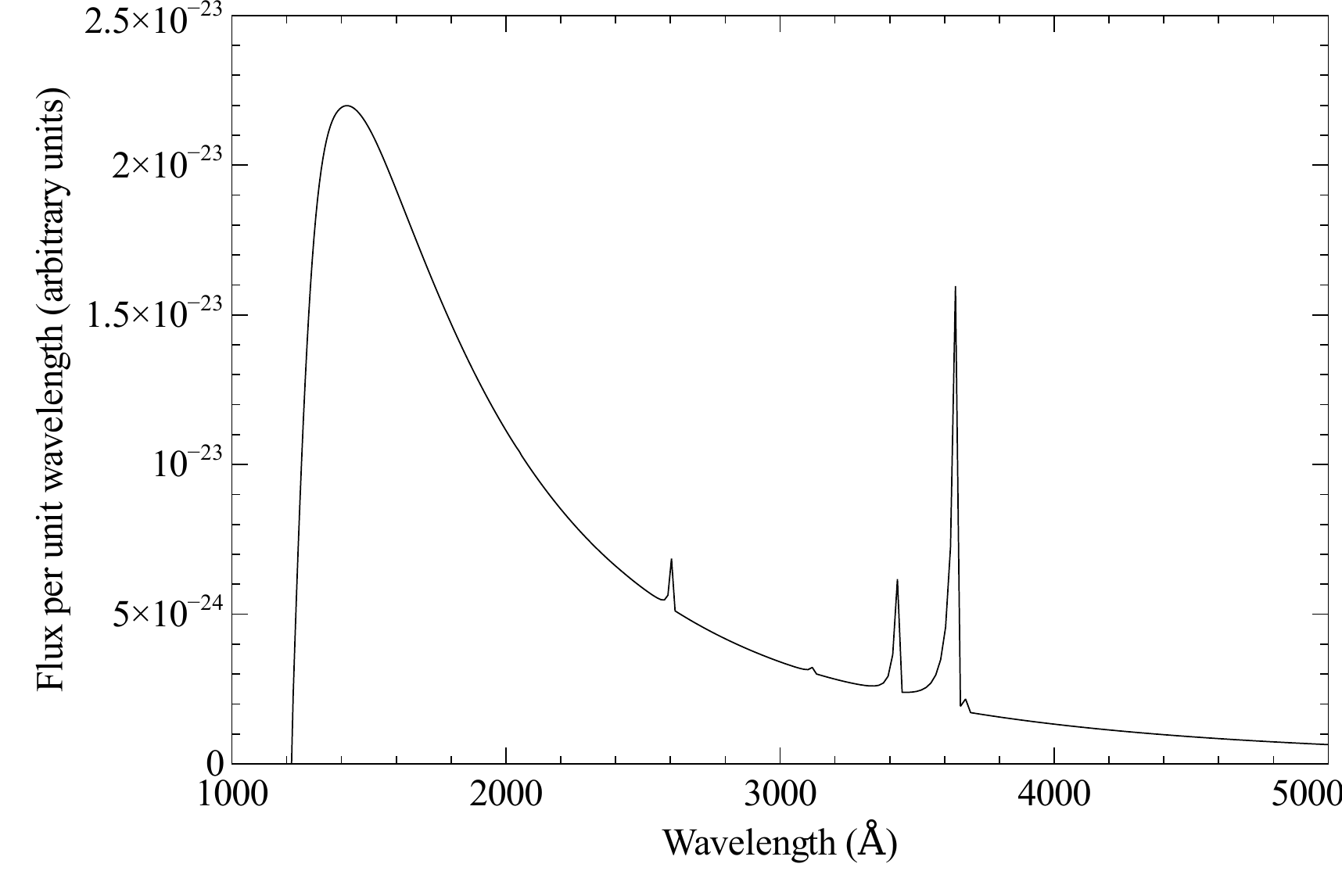}
\caption{Spectrum of continuum emission produced by the particle heating model
with $\alpha$=0.2, and pressure P=$10^{6.5}\pccmK$.}
\label{contspec}
\end{figure}

We note that from Section 3.1 the C~I$\lambda1656$\AA\ line is predicted to be
the strongest emission line in the F140LP band. In our model the flux ratio of
this line to H$\alpha$ is 0.2 so this line contributes less than 4 per cent of
the counts in this band.

\section{Comparing the ratio R=F$_{1500}$/I(H$\alpha$) between model and
observations}
If we ignore any intrinsic reddening within the filament we obtain the ratio of
the far ultraviolet continuum at 1500\AA\ to the H$\alpha$ flux, deduced from
the HST imaging data, a value of R = $4.1\times10^{-3}$\AA$^{-1}$. This value is
just half the value of R=$8.2\times10^{-3}$\AA$^{-1}$  predicted by the particle
heating model. We note here that although the observed fluxes have been
corrected for reddening in our Galaxy, there is also reddening intrinsic to the
filamentary system in NGC~1275. This can be seen by comparing the value of
I(H$\beta$)/I(H$\alpha$)=0.24 (corrected for Galactic reddening) given in Table
5 of \citet{Ferlandetal09} with that predicted by the particle heating model,
0.271 (Table A1 of \citealt{Ferlandetal09}; this value is unchanged by the new
value of $\alpha$ used in this work). Assuming a Galactic extinction law and
using equations 1 and 2 of \citet{Canningetal11}, which make the simplifying
assumption that the intrinsic reddening occurs in a screen in front of the
Horseshoe 
filament, we determine an intrinsic A$_{\rm V}$=0.377. This corresponds to
extinction factors of 1.32 and 2.52 at the wavelengths of H$\alpha$ and 1500\AA\
respectively. Applying this correction to the observed value of R increases it
to $7.8\times10^{-3}$\AA$^{-1}$, in very good agreement with that from the
particle heating model.

We note here that were the \citet{Conseliceetal01} value of the H$\alpha$ to be
correct that the observed value of R corrected for both Galactic and internal
reddening would be R=0.002, whereas the predicted value from the particle heating
model with $\alpha=-0.35$ as published by \citet{Ferlandetal09} is a factor of nine
higher at 0.018. The increase in R from the model comes from the higher weighting given
to the high R region near T=$10^4$K when $\alpha=-0.35$ while the decrease in
observed R comes from the increase in H$\alpha$ flux.

\section{Conclusion}
We have shown that Region 11 in the Horseshoe filament of NGC~1275 is detected
in the far ultraviolet at wavelengths of ~1500\AA\ by the Hubble Space Telescope
ACS/SBC camera. The observed count rate is consistent with that expected from
Hydrogen two-photon emission if the filaments are excited by ionizing particles.
These particles could naturally originate from the cooler phases of the hot
intracluster medium. In this region there is no requirement for any further
components to the far ultraviolet emission from hot stars or emission lines such
as CIV$\lambda1550$ from intermediate temperature ($10^5$K) gas. Our particle
heating model predicts a C~I $\lambda1656$\AA\ emission line at about
0.2$\times$I(H$\alpha$) which is not expected in a nebula spectrum dominated by
recombination processes.

\section{Acknowledgments}
ACF acknowledges support by the Royal Society. RMJ and REAC acknowledge support
from the Science and Technology facilities Council. GJF acknowledges support by
NSF (0908877; 1108928; \& 1109061), NASA (07-ATFP07-0124, 10-ATP10-0053, and
10-ADAP10-0073), JPL (RSA No 1430426), and STScI (HST-AR-12125.01 and
HST-GO-12309). PvH acknowledges support from the Belgian Science Policy office
through the ESA PRODEX program.

We thank Paul Hewett for producing spectral energy distributions from the
magnitudes tabulated for the GSC stars.

Some of the data presented in this paper were obtained from the Multimission
Archive at the Space Telescope Science Institute (MAST). STScI is operated by
the Association of Universities for Research in Astronomy, Inc., under NASA
contract NAS5-26555. Support for MAST for non-HST data is provided by the NASA
Office of Space Science via grant NNX09AF08G and by other grants and contracts.

\bsp

\label{lastpage}
\clearpage
\end{document}